\documentclass [11pt]{article}
\usepackage{amsmath}
\usepackage{amssymb}
\usepackage{amscd}
\usepackage{amsthm}
\usepackage{amsopn}
\usepackage{xspace}
\usepackage{verbatim}
\usepackage{amsmath}
\usepackage{amsfonts}
\usepackage{graphicx}
\usepackage{mdframed}

\usepackage{subfigure}
\usepackage{subfigmat}
\def\bbc#1#2{{\rm \mkern#2mu\vbar\mkern-#2mu#1}}
\def\bbb#1{{\rm     I\mkern-3.5mu      #1}}      \def\bba#1#2{{\rm
#1\mkern-#2mu\fudge #1}}
\def\bb#1{{\count4=`#1 \advance\count4by-64  \ifcase\count4\or\bba
A{11.5}\or
\bbb B\or\bbc C{5}\or\bbb D\or\bbb E\or\bbb F \or\bbc  G{5}\or\bbb
H\or
\bbb I\or\bbc J{3}\or\bbb K\or\bbb  L  \or\bbb  M\or\bbb  N\or\bbc
O{5} \or
\bbb P\or\bbc Q{5}\orrrr\b
bb   R\or\bbc   S{4.2}\or\bba   T{10.5}\or\bbc   U{5}\or      \bba
V{12}\or\bba
W{16.5}\or\bba X{11}\or\bba Y{11.7}\or\bba Z{7.5}\fi}}
\def\rat{\hbox{{\rm Q}\kern-.70em\hbox{{\rm I} } }}

\newtheorem{theorem}{Theorem}
\newtheorem{lemma}{Lemma}[section]

\newtheorem{cor}{Corollary}[section]

\newtheorem{acknowledgment*}{Acknowledgment}

\numberwithin{equation}{section}

\newcommand{\Feps}{F^\eps_{\gamma}}

\newcommand{\be}{\begin{equation}}
\newcommand{\ee}{\end{equation}}
\newcommand{\bd}{\begin{displaymath}}
\newcommand{\ed}{\end{displaymath}}

\newcommand{\eps}{\varepsilon}

\newcommand{\R}{\mathbb R}

\renewcommand{\vec}[1]{\boldsymbol{#1}}

\begin{document}
\Large
\begin{center} {\bf Contact angles of liquid drops subjected to a rough
boundary
}\end{center}
\normalsize
\begin{center} G. Wolansky
\end{center}
\small
\begin{center}
Department of Mathematics, Technion, Haifa 32000, Israel
\end{center}
\normalsize
\vskip .5in
\begin{center} {\bf Abstract}\end{center}
The contact angle of a liquid drop on a rigid surface is determined
by the classical theory of Young-Laplace. For chemically homogeneous surfaces,
this angle is a constant.
\par\noindent
We study the minimal-energy configurations of liquid drops on
rough surfaces. Here the actual angle is still constant for
homogeneous surfaces, but the apparent angle can fluctuate widely.
A limit theorem is introduced for minimal energy configuration,
where the rigid surface converges to a smooth one, but the
roughness parameter is kept constant. It turns out that the limit
of minimal energy configurations correspond to liquid drop on a
smooth surface with an appropriately defined effective chemical
interaction energy. It turns out that the effective chemical interaction depends linearly on the roughness in a certain range of parameters, corresponding to {\em full wetting}. Outside this range the most stable configuration corresponds to a partial wetting and the effective interaction energy depends on the geometry in an essential way.
 This result partially justifies  {\em and extends} Wenzel and Cassie's laws
and can be used to deduce the actual inclination angle in the most stable state,  where the
apparent one is known by measurement. This, in turn, may be
applied to deduce the roughness parameter if the interfacial
energy is known, or visa versa.  \vskip 3in\noindent

 {\bf Key words:} \ \ Liquid drops, mean curvature, Young angle,
 Wenzel angle, functions of bounded variations.
\newpage
\section{Introduction}
The classical theory of the shape of liquid drops is related to
the theory of surfaces with a prescribed mean curvature (PMC). The
beginning of the modern theory of PMC is dated back to the early
18th century, and is known today as the Young-Laplace theory
\cite{[Y]}, \cite{[L]}. A great progress in the understanding of PMC and their rich
 structure was achieved in the second half of the 20th century,
together with the development of BV theory.
and the geometric  measure theory.
In addition, the classical theory of minimal surfaces was
advanced using analytic and topological methods.
\par
A particular aspect of this theory is the inclination angle of the
liquid-solid phases at the intersection line of the
liquid-solid-vapor. This angle attracts a lot of attention in the
physics and chemistry literature because it is determined by the
chemical properties of the liquid and solid phases, and may serve
as a practical  device for the actual measurements of such
parameters for different solids (See, e.g. \cite{[NVZ]}, \cite{[TLB]}, \cite{[CPM]}).
\par
However, the details of the interaction energy at the interaction
line is still controversial. Several corrections were suggested to
the classical Young-Laplace theory in the vicinity of the
interaction line, where the liquid phase is very thin ( \cite{[NC]},
\cite{[CS]}, \cite{[S]}).
\par
On top of this, the geometry of the solid surface itself can
complicate  the understanding of the contact-line formation and
the resulting inclination angle. This aspect is also of practical
interest in the study of porous media wettability. For example,
the energy barrier for nucleation in calcium deposits is strongly
affected by the contact angle in the presence of wetting \cite{[JBGM]}.
See also \cite{[MM]} for the study of contact angle on pore throats
formed by spheres.
\par
The effect of roughness of the solid surface on the contact angle
was studied theoretically by several authors. \cite{[HM]}, \cite{[JD]}, \cite{[SG]}
considered the effect of hysteresis, where the equilibrium contact
angle depends on the formation of the drop. This dependence leads,
in particular, to the concept of advancing and receding angles,
formed by equilibrium configurations after inflation and
depletion, respectively, of the drop on a given rough surface. The
hysteresis phenomena is attributed to the presence of local
minimizers of the energy, compatible with Young law \cite{[M3]}.
\par
It seems, however, that a rigorous understanding of the relation between
the local and apparent inclination angle for rough surfaces is still
missing, even in the context of the classical Young-Laplace theory.
A heuristic argument proposed in the late 40's and early 50's by Wenzel
and others \cite{[WN]}, \cite{[D]}, \cite{[SB]}, \cite{[G]} suggested a way to calculate the relation between
the Young angle and the apparent angle. By this argument, the
apparent inclination angle of the {\it global}
energy minimizer is determined by the mean surface energy of the rough
surface.
\par
In this paper we attempt a rigorous justification of Wenzel rule
and study its limitation. The model we adopt is basically the
classical Young-Laplace theory, leading to liquid-vapor surfaces
of prescribed mean curvature with a constant local inclination
angle given by Young rule. We shall demonstrate now a simple
version of this model.
\par
Assume that the 2-dimensional solid surface is a graph of a
prescribed function $z=w(x,y)$; the liquid domain occupies the
subgraph of a function $u$ above the graph of $w$, i.e the liquid
domain is given by $$ \{x,y,z\} \ \
; \ \ w(x,y) \leq z \leq u(x,y)$$ The mean curvature of the graph
of $u$ (the fluid-vapour interface) is given by $div ({\bf T}u)$,
where ${\bf T}u = \nabla u / \sqrt{1 + |\nabla u|^2}$. The
equation describing the liquid-vapor interface $u$ in the domain
$u>w$ is given by \be div({\bf T} u) = \lambda\label{mc}\ee where the constant $\lambda$  is  the mean curvature determined by the volume  of
the droplet, or $\lambda=0$ in the case of a minimal
surface (soap films). The free boundary condition at the
fluid-solid-vapour interface $u=w$ is given by \be \frac{1 +
\nabla u\cdot \nabla w}{\sqrt{1 + |\nabla u|^2} \sqrt{1 + |\nabla
w|^2}} = -\gamma\label{yang}\ee where $\gamma$ is a physical
parameter for the interaction energy between the liquid and the
solid phases.\footnote{We ignore here the vapour-solid interaction
energy. It can be taken into account by a suitable change in
$\gamma$.}
 This constant determines the  inclination
angle
\be\label{Youngang}\theta_Y = \arccos(-\gamma)\ee between the solid and liquid at
the interface line and  is known as Young's angle \cite{[Y]}.
\par
Eq. (\ref{mc}) and the boundary condition (\ref{yang}) are derived
from the free energy \cite{[WM]} \be {\cal F}(u) = \int\int_{u-w\geq 0}
\left[\sqrt{1 + |\nabla u|^2} +  \gamma
\sqrt{1+|\nabla w|^2}\right]dxdy\label{fe}\ee
under the constraint
 \be  \int\int
(u-w)_+
dxdy=q>0 \ . \label{confe}\ee
 Here the volume $q$ is the conjugate to the mean curvature $\lambda$ ( $\lambda=0$ if there is no volume constraint).
\par
Young \cite{[Y]} stated that, for chemically homogeneous solid surface ($\gamma=const$),
the contact angle is constant along the contact line. In particular, for a flat surface $w\equiv 0$ and $\gamma \in (-1,0]$ (hydrophilic surface) the contact angle is identical to the {\em apparent angle} via (\ref{yang})
\be\theta_{app} :=\arccos\left( 1 + |\nabla u|^2\right)^{-1/2}=\theta_Y
\label{act} \ . \ee
If the surface $z=w$ is rough, as is the case in
practical applications, then the {\it apparent} angle given by (\ref{act})
is very sensitive to $\nabla w$ \cite{[M1]}.

From a
mathematical point of view, the contact angle is a problematic
concept.
\par
Consider, for example, the case
\be\label{Weps}w(x,y) = \eps\omega(x/\eps, y/\eps)\ee
where $\omega$ is a periodic function in both variables. For $\eps$ very
small, the solid interface looks flat. On the other hand, $\nabla w$ is
of order one and the local Young angle (\ref{yang}) may deviate significantly
from the apparent inclination $\theta_{app}$ (\ref{act}).
\par
\begin{figure}
 \centering
\includegraphics[height=5.cm, width=10.cm]{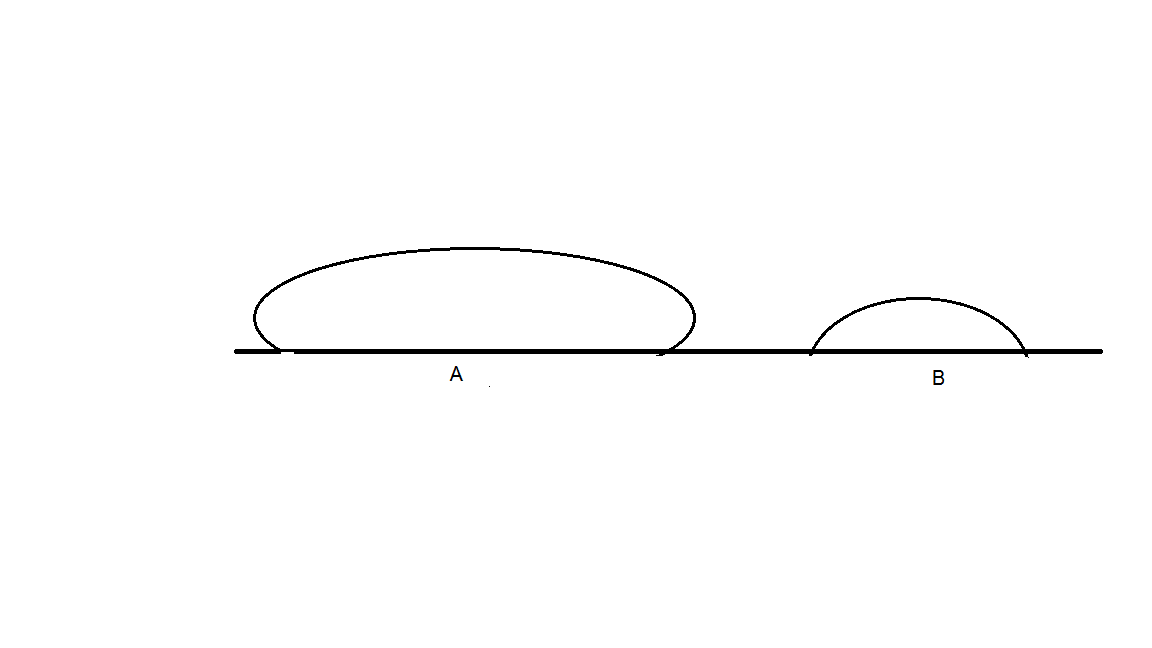}\\
\caption{A: hydrophobic droplet; \ \ B: Hydrophilic droplet}
\label{phobicphilic}
\end{figure}
A natural resolution of this problem is to replace the last term
in the free energy ${\cal F}$ by $\gamma_w:=r\gamma$, where $r\geq 1$ stands for  roughness of the solid surface, measuring the local ratio of its surface area to the  surface area  to its smooth approximation
 \cite{[M2]}. In the particular case (\ref{Weps}) we obtain that $r$ is a constant given by the average of $\sqrt{1+|\nabla w|^2}$ over a period. Thus, we minimize the
"effective" free energy, i.e the free energy on a flat surface
with an effective interaction energy $\gamma_w = r\gamma$: \be
{\cal F}_{eff}(u) = \int\int_{u-w\geq 0}\left[\sqrt{1 + |\nabla u|^2}+ \gamma_w\right]dxdy\label{effe}\ee This yields the
inclination {\it Wenzel angle} $\theta_W$ defined as
 \be\label{Wa} \theta_{W}= \arccos(-r\gamma)\ee, known as "Wenzel
rule" \cite{[WN]}.
\par
In the literature, Wenzel law is usually associated with both {\it complete wetting} and {\it the most stable configuration}  \cite{[CPM]}. The case of incomplete wetting is usually attributed to a meta-stable state and is associated with the Cassie-Baxter equation (Cassie's law) \cite{[AMAR]}
\be\label{casinib} \cos\theta_{app}= \rho f \cos(\theta_Y) +f-1\ee
where $f\in [0,1]$ stands for the fraction of the wetted surface, $\rho$ the roughness parameter in the wetted portion and $\theta_Y$ is the homogeneous Young angle ( in the current notation $\cos(\theta_Y)=-\gamma$).
\par
\begin{mdframed}
To the best of our knowledge, there
is no rigorous justification for the Wenzel rule {\em as a description for the apparent angle of the free energy global minimizer} (e.g. the "most stable state") in the hydrophobic ($\gamma>0$) range.
\end{mdframed}

 In this paper we shall address this problem. We first note that the formulation of the Free energy (\ref{fe}) is not consistent in the
  hydrophobic case , at least for  an {\it approximately flat} surface $w\approx 0$. Indeed, in that case, both Young and Wenzel angles are obtuse, so the liquid phase cannot be obtained as a subgraph of a function $u$ (Fig. \ref{phobicphilic}
 ). In order to handle the  hydrophobic case, we formulate the free energy in terms of an {\em unparameterized} functional. For this, we consider the liquid domain $E$ contained in a {\em bounded container} $\Omega\subset \R^n$ whose boundary $\Gamma$ is assumed to be a smooth, closed $n-1$ dimensional surface.\footnote{ Of course, the physical situation in our world corresponds to $n=3$.}
  The Free energy $F_\gamma=F_\gamma(E)$ is defied as
  $$ F_\gamma(E)=P_\Omega(E)+ T_\Gamma(\gamma E)$$
  where $P_\Omega(E)$ stands for the {\em relative perimeter} of $E$ in $\Omega$ and $T_\Gamma(\gamma E)$ stands for the $\mathbb{L}_1$ norm of the  trace of the {\em function} $\gamma$ (defined on $\Gamma$), on $\Gamma\cap \partial E$. Both notions are reviewed in section \ref{review}. The stable states are the minimizers of $F_\gamma$ under a constraint on the volume  $Vol(E)=q<Vol(\Omega)$.

In this formulation, the Young angle $\theta_Y$ is defined at any point in the boundary of $\Gamma\cap \partial E$ as the angle between the normals to $\Gamma$ and $\partial E$ at this point. Formally, it satisfies definition (\ref{Youngang}).
\par
\begin{figure}
 \centering
\includegraphics[height=5.cm, width=10.cm]{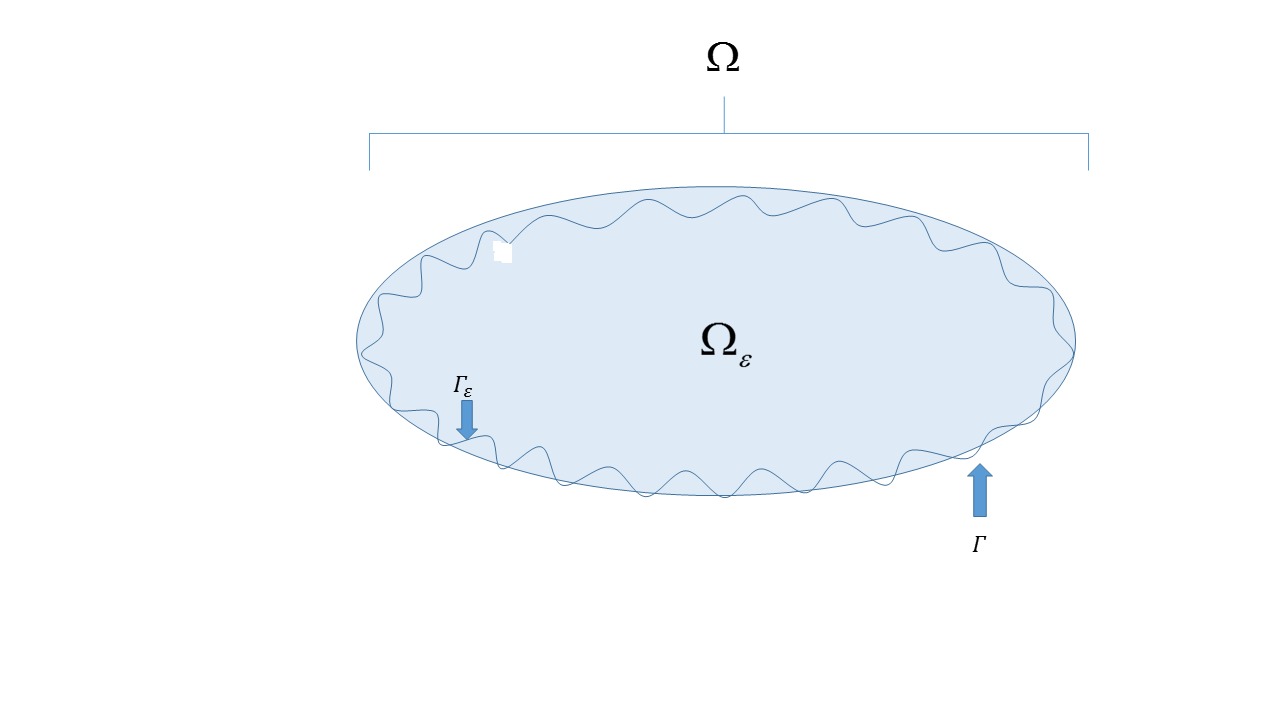}\\
\caption{Rough approximation of $\Omega$}
\end{figure}
In this paper we consider a family of "rough domains" $\Omega_\eps\subset \Omega$ which approximate $\Omega$ in the sense $\Omega_\eps\rightarrow \Omega$ as $\eps\rightarrow 0$. The roughness parameter of this family is defined, naturally, as a function $r$ on $\Gamma$ satisfying the trace limit
\be\label{Rdef} T_\Gamma (r\phi)=\lim_{\eps\rightarrow 0} T_{\Gamma_\eps}(\phi) \ ,  \ee
where $\Gamma_\eps$ the boundary of $\Omega_\eps$,
for {\em any} smooth function $\phi$ defined on the closure $\Omega\cup\Gamma$.
\vskip .3in\noindent
\begin{mdframed}
The {\em apparent contact angle} for the rough approximations $\{\Omega_\eps\}$ is {\bf defined} as
$$\theta_{eff}=\arccos(-\gamma_{eff}) \ , $$ provided $F_{\gamma_{eff}}$ is the limit of the functionals $F_\gamma$ on $\Omega_\eps$. By this we mean:
\par
{\it If $E_\eps\subset \Omega_\eps$ is a minimizer of
$$F_\eps(E):= P_{\Omega_\eps}(E)+ T_{\Gamma_\eps} (\gamma E)$$
subject to the constraint $Vol(E)=q$, then there exists a limit $\bar{E}=\lim_{\eps\rightarrow 0}E_\eps\subset\Omega$ which is a minimizer of
$$ F_{\gamma_{eff}}(E):= P_\Gamma(E) + T_{\Gamma}(\gamma_{eff} E)$$
subjected to the same volume constraint. }
\end{mdframed}
  The Wenzel rule is, then, justified if $\gamma_{eff}=r\gamma$, were $r$ as defined in (\ref{Rdef}).
\par
It is conceivable that the Wenzel rule is satisfied for global minimizers in the hydrophilic case $\gamma<0$, so we concentrate in the case $\gamma>0$. For simplicity we assume that $\gamma$ is a constant on $\Gamma$.
\par
We further assume that $\gamma$ is a constant. Our first result is:

\begin{center} There exists a critical $0<\gamma_c<1$ such that $\gamma_{eff}=r\gamma$ if $\gamma\leq \gamma_c$.
\end{center}
The definition of $\gamma_c$ is given by (A4) or (A'4) in section \ref{rough}. In particular
\begin{mdframed} The validity of Wenzel rule in the hydrophobic case is guaranteed only for $\gamma\leq\gamma_c$.
\end{mdframed}
The post critical case $1\geq \gamma>\gamma_c$ is discussed in Section \ref{pw}. For simplicity, we concentrate on the two-dimensional case where the boundary of $\Omega_\eps$ looks, locally, as a graph of a period-$1$ extension of an {\em even}   function $\zeta=\zeta(s)$ on $[-1/2,1/2]$, on the $\eps$ scale. The main result introduced in Theorem \ref{pw}.1 yields the existence of an explicit function which, under certain generic assumptions on $\zeta$, takes the form
\be\label{gammaeff} \gamma_{eff}(\gamma):= 2\min_{0\leq s\leq 1}\left[ s+\gamma\int_s^{1/2}\sqrt{1+|d\zeta/ds|^2}ds\right] \ . \ee
Recalling that the roughness $r$ consistent with (\ref{Rdef}) is given, in the above case, by $$r=2\int_0^{1/2}\sqrt{1+|\zeta^{'}|^2}ds \ , $$
 we obtain that
$\gamma_{eff}(\gamma)<r\gamma$ in this range. This function indicates  {\em deterministic values} for the wetted parameters $f$ and the roughness $\rho$ in the Cassie rule (\ref{casinib}) {\em as functions of $\gamma$}:
$$ f=1-2s_0\ \ \ ; \ \ \ \rho=\frac{\int_{s_0}^{1/2}\sqrt{1+|\zeta^{'}|^2}ds}{1/2-s_0}$$
where $s_0\in (0, 1/2)$ is the minimizer of (\ref{gammaeff}). In particular

\begin{mdframed}
The Cassie rule with the prescribed parameters represents the {\em most stable} droplet configuration in the post critical interface energy $\gamma>\gamma_c$.
\end{mdframed}
\subsection{{\it Layout}}
Our approach to this problem is via the theory of $BV-$sets \cite{[G1]}.
 In  section~\ref{review} we review the free energy functional in
this setting, and some basic facts on the $BV-$space. In
section~\ref{rough} we describe some assumptions on the rough
domains. In section~\ref{main} we collect some auxiliary results
which, in general, are well known, but not necessarily in the form
we introduce. The main results of this paper are given in sections
\ref{proofmain} (full wetting) and \ref{pw} (partial wetting).
\section{The free energy for capillary surfaces: A review}\label{review}
{\bf Notations and standing assumptions:}\par\noindent
\begin{description}
\item{i.} \ A {\it cavity} is a bounded domain $\Omega\subset
\R^n$ which contains the fluid and vapour phases.
\item{ii} \ The {\it volume} (Lebesgue measure) of $\Omega$ is $V>0$,
\item{iii.} \ The interface  of the solid phase with the
fluid/vapour phases is the boundary of $\Omega$, denoted by
$\Gamma$. The closure of $\Omega$ is $\Omega \cup\Gamma \equiv
\Omega^c$. We shall always assume that $\Gamma$ is, at least, a
Lipschitz surface.
\item{iv.} \ ${\cal H}_{n-1}$ is the $n-1$ dimensional   Hausdorff
measure on $\Gamma$
\item{v.} \ $\vec{n}$ is the
outward normal to $\Gamma$ pointing into the complement of
$\Omega$. For a Lipshitz surface $\Gamma$,  $\vec{n}$  is defined
for ${\cal H}_{n-1}-$ almost any point $x\in\Gamma$.
\item{vi.} \ The fluid-solid interface energy is a continuous
function $\gamma$ defined on $\Gamma$. It is assumed that
$0\leq\gamma\leq 1$ on $\Gamma$ (We may also be assumed that $-1
\leq \gamma\leq 0$. We shall take the case of nonnegative $\gamma$
but there is no limitation of generality).
\item{vii.} \ The cavity $\Omega$ is said to be {\it smooth} if
there exists a vector-field $\vec{v}\in C^1(\Omega^c \ ; \ \R^n)$
such that $\left|\vec{v}(x)\right|\leq 1$ for any $x\in\Omega^c$
and $\vec{v} = \vec{n}$ a.e on  $\Gamma$.
\item{viii.} \ A set $E\subset\Omega$ is the fluid domain.FIn particular
$\phi_E$ is the characteristic function corresponding to $E$ in
$\Omega$, i.e $\phi(x) = 1$ if $x\in E$, $\phi(x)=0$ if
$x\in\Omega - E$. 
\item{ix.} \ A function $\phi\in \bb L_1(\Omega)$ is of bounded
variation in $\Omega$ if $\int_\Omega|\nabla\phi| < \infty$ where
$$  \int_\Omega |\nabla\phi|\equiv \sup_w\left\{ \int_\Omega \phi
div(\vec{w}) \ ; \ \ \vec{w}\in C^\infty_0 (\Omega; \R^n) \ \ , \
\ |\vec{w}|_\infty \leq 1\right\}$$
  The space of functions of bounded variation in
$\Omega$ is $BV(\Omega)$. The $BV-$norm  is $||\phi||_{BV} \equiv
\int_\Omega |\nabla\phi| + |\phi|_1$ where $|\phi|_1:= \int_\Omega
|\phi|$.
\item{x.} The  {\it
perimeter} of a set $E$ in $\Omega$ is  $P_\Omega(E):=\int_\Omega
|\nabla\phi_E|$. A set $E$ of finite perimeter is called a
 Caccioppoli  set. The collection of  Caccioppoli sets $E\subset
\Omega$ of a prescribed volume $Vol(E):=|\phi_E|_1=q$, $0 < q < V$ is
denoted by $\Lambda_q$. We shall use sometimes use $E\in
BV(\Omega)$ for a Caccioppoli set.
\item{xi.} \ The {\it Free-Energy} corresponding to a function
$\phi\in BV(\Omega)$ is
$$F_\gamma(\phi) = \int_\Omega |\nabla\phi|
+ \int_\Gamma \gamma\phi d{\cal H}_{n-1}$$
We shall also refer to
$F_\gamma(E)= F_\gamma(\phi_E)$.
\end{description}

  \par\noindent It is
known \cite{[G1]} that for any Lipschitz surface
$S\subset\overline{\Omega}$, the trace of a BV function on $S$ is
defined in $\bb L_1(S)$. In particular, the trace of a Caccioppoli
set $E$ is defined on $S$. Moreover, $\left.\phi_E\right|_S \in
\bb L_\infty(S)$ and $0 \leq \phi_E \leq 1$  a.e on $S$.

\par
We recall  the compactness property of $BV$ functions \cite{[G1]}: \vskip
.2in\noindent{\bf Compactness:} \ {\it A sequence $\phi_j\in
BV(\Omega)$ bounded uniformly in the $BV$ norm contains an $\bb
L_1-$ converging subsequence to some $\phi\in BV(\Omega)$.
Moreover, $\int_\Omega \left|\nabla \phi\right|  \leq
\liminf_{j\rightarrow\infty} \int_\Omega\left|\nabla \phi_j\right|
$. If $\phi_j$ are characteristic functions of Caccioppoli sets
$E_j$, then any limit $\phi$ is also a Caccioppoli set
$E\subset\Omega$. } \vskip .2in The compactness theorem clearly
yields the existence of a minimizer to $F_0$ ($\gamma=0$). If
$\gamma\not= 0$ then the trace of a sequence of a $BV$ sets is to
be taken into account. It can be easily shown that this trace is
neither upper semi-continuous, nor lower semi-continuous in the
underlying space. To handle the trace,  the following perimetric
inequality is applied  (\cite{[E]}, see also Lemma 6.1 in \cite{[F]})

\vskip .2in\noindent {\bf Lemma~\ref{review}.1}: If L is the
minimal Lipschitz constant of $\Gamma$ then for any $\delta > 0$
we may choose $C=1+L + \delta$ and a corresponding
$\beta=\beta(\delta)$ for which \be\int_\Gamma |\phi| d{\cal
H}_{n-1} \leq C\int_\Omega\left|\nabla \phi\right|
 + \beta|\phi|_1\label{pe}\ee
holds for any $\phi\in BV(\Omega)$.
\par
Using the perimetric inequality (\ref{pe}) and the compactness of
$BV$ space it is possible to prove the existence of a minimizer to
$F_\gamma$ in $\Lambda_q$ for $|\gamma|$ small enough. The
following theorem is a slight generalization of Theorem~1.2 in
\cite{[G2]}:

\begin{theorem}\label{review1} If the perimetric inequality (\ref{pe})
holds for  $C\leq 1/|\gamma|$ then there exists a minimizer $E_0$
of $F_\gamma$ in $\Lambda_q$ for any $0<q< V$. \end{theorem}

The main step for the proof of Theorem \ref{review1}
is the inequality $$
\int_\Gamma \gamma | \phi|
< \int_\Omega|\nabla\phi| + \beta^{'}|\phi|_1$$ which follows from
(\ref{pe}) together with the assumptions of the theorem. This
yields, essentially, that $F_\gamma$ is lower-semi-continuous in
the underlying spaces.
\par
If $\Omega$ is smooth (in the sense of
notation vii), then the perimetric inequality (\ref{pe}) can be
replaced by \be\int_\Gamma |\phi| d{\cal H}_{n-1} \leq
\int_\Omega\left|\nabla \phi\right|
 + \beta|\phi|_1\label{spe}\ee
for some $\beta > 0$. Hence Theorem~\ref{review}.1 implies, for a
smooth domain $\Omega$, the existence and smoothness of a
minimizer for $|\gamma|\leq 1$ (i.e for any inclination angle
$-\pi\leq\theta\leq \pi$). The inequality (\ref{spe}) seems to be
known to experts, but we did not find a proof for it in the
literature. For completeness, we will introduce the  proof of
(\ref{spe}) as a part of a more general result in section~4.

\section{Rough  domains}\label{rough}
Let us now consider a rough domain $\Omega_\eps$. We shall adopt
the notation $i-xi$ of section~2 for the domain $\Omega_\eps$,
adding the index $\eps$. Thus, $\Gamma_\eps$ is the boundary of
$\Omega_\eps$, $\vec{n}_\eps$ is the outward normal to this
boundary, etc. Below we pose our assumptions on the perturbed
domain.
\begin{description}
\item{A1.}
For every $\eps > 0$,
$\Omega_\eps \subset \Omega$ is a Lipschitz domain.
\item{A2.} \ \
$\lim_{\eps\rightarrow 0} \Omega_\eps = \Omega$
\end{description}
\noindent Our results on partial wetting (section~6) require us to
allow the solid-liquid interaction to depend on $\eps$, that is
$\gamma = \gamma_\eps(x)$ is a function defined on $x\in\Gamma$ and $\eps$. We
further assume:
\begin{description}
\item{A3.} 
 There exists
$\gamma_w\in \bbb L_\infty (\Gamma)$ such that for any
$\phi\in BV(\Omega\cup\Gamma)$,  $$\lim_{\eps\rightarrow
0}\int_{\Gamma_\eps}\gamma_\eps\phi d{\cal H}_{n-1}   = \int_\Gamma
\gamma_w\phi d{\cal H}_{n-1}$$
\item{A4.} The domain $\Omega$ is smooth (see vii, section~2).
Let $\vec{v}$ be the vector-field defined in (vii). Then
\be\label{defgamacx} \gamma_\eps(x)\leq \gamma_c:= \liminf_{\eps\rightarrow 0}\inf_{x\in\Gamma_\eps}\vec{n}_\eps(x)\cdot\vec{v}(x) \ . \ee
\end{description}
In particular $\gamma_\eps(x)\leq 1$ for any $x\in\Gamma_\eps$.
If $\gamma_\eps=\gamma$ is a constant in both $\eps$ and $x$  and
$\Omega$, $\Omega_\eps$ are smooth domains, then we may replace
assumptions (A3, A4) by:
\begin{description}
\item{A'3.} \
Let $B(x,\delta)$ be the ball of radius
$\delta$ centered at $x$.
Then
there exists a function $r\in \bbb L_\infty(\Gamma)$ such that
$$ \lim_{\delta\rightarrow 0}\lim_{\eps\rightarrow 0}
\frac{{\cal H}_{n-1}(\Gamma\cap B(\delta, x))}
{{\cal H}_{n-1}(\Gamma\cap B(\delta, x))}
= r(x)$$
holds uniformly on $\Gamma$, and $\gamma_w=\gamma r(x)$.
\item{A'4.} \
\be\label{gammacconst}\gamma \leq \lim_{\delta\rightarrow 0}\liminf_{\eps\rightarrow
0}\inf\left\{ \vec{n}(x)\cdot \vec{n}_\eps(y) \ \ ;  \ \ x\in\Gamma, \  \ \ y\in
\Gamma_\eps\cap B(\delta, x)\right\}:=\gamma_c\ . \ee
\end{description}
\noindent {\bf Remark:} The number $r(x)$ in A'.3 is  the local
roughness parameter  \cite{[M2]}. \vskip .2in\noindent {\bf
Proposition~\ref{rough}.2} \ {\it If $\gamma\geq 0$ is a constant
(independent of $\eps$)  and $\Omega$, $\Omega_\eps$ are smooth
domains, then conditions A1, A'3 and A'4 imply  A3 and A4 where
$\gamma_w = \gamma r$.}
\par
\par\noindent{\bf Proof:}
Let $x\in \Gamma$. We may assume that in the neighborhood of $x$,
$\Gamma$ can be described locally as a graph of a $C^1$ function
$x_n= \psi(x^{'})$ where $x^{'} = (x_1, \ldots x_{n-1})$. We may
further assume that $x^{'} = 0$, hence $x= (0,\psi(0))$, while
$\nabla\psi(0) = 0$. Since $\gamma_c > 0$ by assumption, it follows
that, for a sufficiently small $\eps > 0$ and in a sufficiently
small neighborhood of $x$, the section of $\Gamma$
intersecting this neighborhood is also a graph of a function $x_n
= \psi_\eps(x^{'})$. Since $\vec{n}_0 = \{ \vec{0} , 1\}$ at the
point $x$ and $\vec{n}_\eps = ( 1 + |\nabla\psi_\eps|^2)^{-1/2}
\left( -\nabla\psi_\eps, 1\right)$, we obtain by A'.4:
$$\frac{1}{(1 + |\nabla\psi_\eps(x^{'})|^2)^{1/2}} \geq
\gamma_c(x)$$ for a sufficiently small $\eps$ in a sufficiently
small neighborhood of $x^{'} = 0$. On the other hand $$r(x) =
\lim_{\delta\rightarrow 0} \lim_{\eps\rightarrow 0}
\frac{\int_{|x^{'}| \leq \delta}\sqrt{1 + |\nabla\psi_\eps|^2}}{
\int_{|x^{'}| \leq \delta}\sqrt{1 + |\nabla\psi|^2 }} =
\frac{1}{B_{n-1}}\lim_{\delta\rightarrow 0}\delta^{1-n}
\lim_{\eps\rightarrow 0} \int_{|x^{'}| \leq \delta}\sqrt{1 +
|\nabla\psi_\eps|^2}$$ where $B_{n-1}$ is the volume of the $n-1$
unit ball. This implies that $r$ is, in fact, the local average of
$\vec{n}_0(x)\cdot \vec{n}_\eps(y)$ and the inequality $r <
1/\gamma_c$ follows.
\par
To complete the proof we show, under the above condition,
$$\lim_{\eps \rightarrow 0}\int_{\Gamma}\phi d {\cal H}_{n-1}
 = \int_\Gamma r\phi d {\cal H}_{n-1}$$
for any  $\phi\in BV(\Omega)$.
Following the same line as above, we obtain that in a neighborhood
$B\subset \Gamma$
($B_\eps\subset \Gamma$)
given by the graph of $\psi$ ($\psi_\eps$) over a set $D\subset\R^{n-1}$
$$\int_{B\cap\Gamma} \phi = \int_{D_\eps}
\sqrt{1 + |\nabla\psi_\eps|^2}\phi\left(x^{'}, \psi_\eps(x^{'}\right)
dx^{'}$$
\be = \int_{D_\eps}
\sqrt{1 + |\nabla\psi_\eps|^2}\phi\left(x^{'}, \psi(x^{'})\right) dx^{'}
+ \int_{D_\eps}
\sqrt{1 + |\nabla\psi_\eps|^2}\left[\phi\left(x^{'}, \psi_\eps(x^{'})\right)
- \phi\left(x^{'}, \psi(x^{'}\right)\right] dx^{'}\label{pro2}\ee
Let $\delta(\eps)$ be the distance between $\Gamma$ and $\Gamma$, and
$\Omega_\delta = \{ x\in\Omega \ ; \ dist(x, \Gamma) < \delta\}$.
\par\noindent
The second term in (\ref{pro2}) is estimated by
$\int_{\Omega_\delta}|\nabla\phi|\rightarrow 0$ as $\eps\rightarrow 0$.
Since $\phi\left(x^{'}, \psi(x^{'})\right)\in \bb L_1(D)$ we obtain the
convergence of the first part of (\ref{pro2}) to $\int_D r(x^{'})\phi
\left(x^{'},\psi(x^{'})\right)$. \ \ \ \ \ \ \ \ $\Box$
\vskip .2in\noindent
{\bf Example:} \ Let $\Omega$ be given by a supergraph of a
 function $x_n> w(x^{'})$, and let
$\Omega_\eps = \left\{ x_n > w_\eps(x^{'})\right\}$
 where
$w_\eps(x^{'}) = w(x^{'}) + \eps\zeta(x^{'}/\eps)$,  while
$\zeta > 0$ is a periodic function on the torus
$[0, 2\pi]^{n-1}$.
\par\noindent
Then condition A'3 holds with
$$r(x^{'})= \frac{\int_{[0, 2\pi]^{n-1}}\sqrt{1 + \left|\nabla w(x^{'}) +
\nabla_{q}\zeta(q)\right|^2}d^{n-1}q}{(2\pi)^{n-1}\sqrt{1 +
|\nabla w(x^{'})|^2}}$$
where
$$\gamma_c =\inf_{x\in \R^{n-1}}\inf_{y\in [0,\pi]^{n-1}}
\frac{1 + | \nabla w(x)|^2 + \nabla w(x)\cdot \nabla \zeta(y)}{(1 +
|\nabla w(x)|^2)^{1/2}(1+|\nabla w(x) + \nabla\zeta(y)|^2)^{1/2}}
$$
\section{Auxiliary results}\label{secmain}\label{main}
The key parametric inequality (\ref{pe}) for
Theorem~\ref{review1}
 can be found in \cite{[F]}, p. 142,  using a partition of the boundary
$\Gamma$ and direct estimates on the trace of $\phi$. In the case
of a smooth domain $\Omega$ there is an alternative way to prove
the stronger inequality (\ref{spe}), using an extension of Gauss
Theorem to $BV$ functions. It follows that, for any vector field
$\vec{v}\in C^1(\Omega^c)$ \be \int_\Omega \phi \nabla\cdot\vec{v}
= -\int_\Omega\vec{v}\cdot\nabla\phi + \int_\Gamma
\phi(\vec{v}\cdot \vec{n}) d{\cal H}_{n-1}\label{gauss}\ee holds
for $\phi\in BV(\Omega)$, where the R.H.S is  defined since
$\nabla\phi$ is a vector-valued Radon measure and $\phi|_\gamma\in
\bb L_1(\Gamma)$. Moreover, (\ref{gauss}) holds for Lipschitz
domains $\Omega$ as well, were the normal $\vec{n}$ to $\Gamma$ is
defined a.e.
\par
Another  item  which we need is the   coarea formula \cite{[G1]}: $$
\int_\Omega |\nabla \phi| = \int_{-\infty}^\infty
dt\int_\Omega\left|\nabla\phi_{F_t}\right|$$ where $F_t = \{
x\in\Omega \ ; \ \phi(x) < t\}$. This leads, in particular, to
$$\int_\Omega |\nabla\phi_+| = \int_0^\infty
dt\int_\Omega\left|\nabla\phi_{F_t}\right| \ \ \ ; \ \ \
\int_\Omega |\nabla\phi_-| = \int_{-\infty}^0
dt\int_\Omega\left|\nabla\phi_{F_t}\right|$$ where $\phi_\pm$ is
the positive/negative part of $\phi$. Since $|\phi| = \phi_+ +
\phi_-$, this leads, in particular, to the conclusion that
$|\phi|\in BV(\Omega)$ if $\phi\in BV(\Omega)$ and
\be\int_\Omega|\nabla|\phi|| =
\int_\Omega|\nabla\phi|\label{abs}\ee In the case of a smooth
domain, we may substitute the vector-field $\vec{v}$ (vii,
section~2) in (\ref{gauss}) to obtain $$\int_\Gamma\phi d{\cal
H}_{n-1} = \int_\Gamma\phi(\vec{v}\cdot \vec{n}) \leq
\int_\Omega|\nabla\phi| + \beta^{'}|\phi|_1$$ where $\beta^{'} =
|\nabla\cdot \vec{v}|_\infty$.  Splitting $\phi$ into its positive
and negative parts and using (\ref{abs}) we obtain (\ref{spe}).
\par
Let us now define, analogously to (xi, section~2), the Free-Energy
of the perturbed domain
\be\label{Feps} \Feps(E) =
\int_{\Omega_\eps} |\nabla\phi_E| + \int_{\Gamma} \gamma
\phi_E d{\cal H}_{n-1}\ee where $E\subset\Omega_\eps$.
\par
Our object is to show that, under assumptions A1-A4, there exists
a minimizer $E_\eps\in\Lambda^\eps_q$ of $F^\eps_{\gamma}$,
where $$\Lambda^\eps_q = \left\{ E\in BV(\Omega_\eps), \ \ vol(E)
= q\right\}$$
This result is not implied directly from
Theorem~\ref{review1} and (\ref{spe}), since $\Omega_\eps$ are
only Lipshitz domains by assumption A.1. On the other hand, we
shall obtain the existence of such a minimizer provided
(\ref{spe}) is replaced by \be\int_{\Gamma} \left|\gamma
\phi\right| d{\cal H}_{n-1} \leq \int_{\Omega_\eps}|\nabla\phi| +
\beta^{'}|\phi|_1\label{sspe}\ee for any nonnegative $\phi\in
BV(\Omega_\eps)$. The inequality (\ref{sspe}) follows, again, by
substituting $\vec{v}$ given by (vii, section~2) in (\ref{gauss}),
obtaining for any nonnegative $\phi\in BV(\Omega_\eps)$ \be
\int_{\Gamma} \gamma\phi d{\cal H}_{n-1} \leq
\int_{\Gamma} \left(\vec{v}\cdot \vec{n}\right)\phi  d{\cal
H}_{n-1} \leq \int_{\Omega_\eps} |\nabla\phi| +
\left|\nabla\cdot\vec{v}\right|_\infty|\phi|_1\label{ssspe}\ee
where we used the assumption $\gamma \geq 0$ and  A4.
Inequality (\ref{sspe}) follows from (\ref{ssspe}) using, again,
the splitting of $\phi$ into its positive and negative parts and
an application of (\ref{abs}). In addition, we obtain that the
constant $\beta^{'}$ in (\ref{sspe}) is  {\it independent} of
$\eps$. This will be crucial in section~\ref{proofmain}.
\par
We shall also need the following results whose proofs follow
directly from definition.
\par\noindent
Let us consider a splitting of a domain $\Omega$ into a pair of subdomains
$\Omega_1$ and $\Omega_2$ such that
\begin{description}
\item{a.} \ $\Omega_1\cap \Omega_2 = \emptyset$
\item{b.} \ $\Omega^c = \Omega_1^c\cup \Omega_2^c$
\item{c.} \ $\Omega_1^c \cap \Omega_2^c \equiv \Gamma_{1,2}$
is a Lipschitz surface.
\end{description}
Then
\par\noindent{\bf Lemma~\ref{main}.1}: \  Given a function
$\phi\in BV(\Omega)$, define $\phi_i$ the
restriction of $\phi$ to $\Omega_i$ where $i=1,2$. Then
$\phi_i \in BV(\Omega_i)$ and, in particular, the traces of $\phi_i$
on $\Gamma_{1,2}$ are defined in $\bbb L_1(\Gamma_{1,2})$. In addition:
$$ \int_{\Omega} |\nabla\phi| = \int_{\Omega_1} |\nabla\phi_1| +
\int_{\Omega_2} |\nabla\phi_2| +
\int_{\Gamma_{1,2}} |\phi_1 - \phi_2| d {\cal H}_{n-1}$$
In particular, it follows that
\be \int_{\Omega} |\nabla\phi| \geq \int_{\Omega_1} |\nabla\phi_1| +
\int_{\Omega_2} |\nabla\phi_2|\label{sest}\ee
\par\noindent
\begin{lemma}\label{main.2}
Let $$\Omega_\delta = \left\{ x\in\Omega \ \ ; \ \ dist(x, \Gamma)
\geq \delta\right\}$$ where $\Omega$ is, again, a smooth domain.
Let a subdomain $D\subset \Omega$. For any $BV$ set $E\subset D$
we have $$\lim_{\delta\rightarrow 0} \int_{D\cap \Omega_\delta}
\left| \nabla \phi_{E\cap\Omega_\delta}\right| = \int_{D} \left|
\nabla \phi_{E}\right|$$
\end{lemma}
 \vskip .2in\noindent In the rest of the
paper we shall abbreviate $\int_{D\cap \Omega_\delta} \left|
\nabla \phi_{E}\right|:=\int_{D\cap \Omega_\delta} \left| \nabla
\phi_{E\cap\Omega_\delta}\right|$, i.e the restriction of $\phi_E$
to the subdomain $\Omega_\delta$ is understood for the integral.
From Lemma~\ref{main}.1 and \ref{main}.2 we have, in particular
\be \lim_{\delta\rightarrow 0}\Delta(\delta, E)=0 \ \ \
\mbox{where} \ \  \ \Delta(\delta, E):= \int_{D} \left| \nabla
\phi_{E}\right|- \int_{D\cap \Omega_\delta} \left| \nabla
\phi_{E}\right| - \int_{D - \Omega_\delta} \left| \nabla
\phi_{E}\right|\label{Delta}\ee \noindent
 We are now in a position
to prove Theorem \ref{review1}: \\
{\bf Proof:} \ We need only to show the lower-semi-continuity of
$F_{\gamma}$. Following \cite{[G2]}, we let $\delta > 0$ and
define $\Omega_\delta$ as in Lemma~\ref{main}.2. Let
$E^n\in\Lambda_q$ be a minimizing sequence of
$F_{\gamma}$, converging to $E_0$. Using
lemma~\ref{main}.2 with $D=\Omega$ and $E=E_0$ and
(\ref{sest}) we get $$ F_{\gamma}(E^n) -
F_{\gamma} (E_0) \geq \left(\int_{\Omega_\delta}
\left|\nabla\phi_{E^n}\right| - \int_{\Omega_\delta}
\left|\nabla\phi_{E_0}\right|\right) + \left( \int_{\Omega
-\Omega_\delta} \left|\nabla\phi_{E^n}\right| -
\int_{\Omega-\Omega_\delta} \left|\nabla\phi_{E_0}\right|
\right)$$ $$ - \int_{\Gamma} \gamma\left|
\phi_{E^n}- \phi_{E_0}\right|d{\cal H}_{n-1}- \Delta(\delta,
E_0) \equiv (1)_n + (2)_n - (3)_n -\Delta(\delta, E_0) $$ Using
(\ref{sspe}) with respect to the domain
$\Omega-\Omega_\delta$ or, if $\Omega$ is a smooth
domain, use $\gamma\leq 1$ to obtain $$ (3)_n \leq
\int_{\Omega-\Omega_\delta}\left|
\nabla\phi_{E^n}\right| +
\int_{\Omega-\Omega_\delta}\left| \nabla\phi_{E_0}\right| +
\beta(\delta) \int_{\Omega_\delta -
\Omega}\left|\phi_{E^n} - \phi_{E_0}\right| \  , $$
hence \be F_{\gamma}(E^n) - F_\gamma(E_0) \geq (1)_n -
2\int_{\Omega-\Omega_\delta}\left|\nabla\phi_{E_0}\right| -
\int_{\Omega_\delta - \Omega}\left|\phi_{E^n} -
\phi_{E_0}\right|- \Delta(\delta, E_0) \ \ .  \label{dde}\ee Now
we let $n\rightarrow \infty$. By the compactness Theorem
(section~\ref{review}) we obtain $\liminf_{n\rightarrow \infty}
(1)_n \geq 0$ as well as the convergence to zero of the third term
on the right of (\ref{dde}). Since $\delta$ is as small as we
wish, the second term  on the right of (\ref{dde}) can also be
made as small as we wish by Lemma~\ref{main}.2, while the last
term goes to 0 by (\ref{Delta}). This implies the
lower-semi-continuity of $F_\gamma$  and the existence of a global
minimizer. \ \ \ \ \ \ \ $\Box$

\section{Complete wetting}\label{proofmain}
Theorem~\ref{review1}, together with the smoothness assumption on $\Omega$  implies the existence of a
minimizer to $F^\eps_{\gamma_\eps}$  in $\Lambda^\eps_q$, for any $\gamma_\eps\leq 1$. Denote such a minimizer by $E_\eps$. For the same reason,  if $\gamma_w\leq 1$ as well, there exists a minimizer $E_0$ of $F_{\gamma_w}$ on $\Lambda_q$.
\par
We now pose our main result: \vskip .3in\noindent
\begin{theorem}\label{proofmain.1}  If $\Omega$ is a smooth domain (vii,
section~2) and $\{\Omega_\eps, \gamma_\eps\}$ satisfy assumptions (A1-A4), then
there exists a subsequence $\eps_n\rightarrow 0$ such that $E_{\eps_n}$ converge in $\mathbb{L}_1(\Omega)$ to $E_0$.
\end{theorem}
 \vskip.2in\noindent

 \begin{cor}\label{proofmain.2} \ Assume (A1-A4)
and, assume, in addition, that the minimum of $F_{\gamma_w}$
is obtained at a {\em unique} set $E_0$. Then the limit
$$\lim_{\eps\rightarrow 0} E_\eps = E_0$$ exists for any choice of a minimize
$E_\eps$ of $F^\eps_{\gamma_\eps}$ in $\Lambda^\eps_q$.
\end{cor}
\vskip .3in
For the proof
of Theorem~\ref{proofmain.1} we will use an elementary version of
the method of $\Gamma-$convergence. In our case, it takes the
following form:
\begin{lemma}\label{proofmain.3}
\ ; \ $\Gamma-$ convergence:  Let $E_0\in\Lambda_q$. Suppose:
\par\noindent
a) \ For any sequence  $E_\eps\in \Lambda^\eps_q$ which converge
in measure to $E_0$, $$ \liminf_{\eps\to
0}F^\eps_{\gamma}(E_\eps) \geq F_{\gamma_w} (E_0)$$
\par\noindent
b) \ There exists such a {\em recovery sequence}  $E^{'}_\eps\in \Lambda^\eps_q$
which converges in measure to $E_0$ and $$\lim_{\eps\rightarrow 0}
F_{\gamma}^\eps (E^{'}_\eps) = F_{\gamma_w}(E_0)$$
\par
Then, any converging subsequence of minimizers of
$F^\eps_{\gamma}$ in $\Lambda^\eps_q$ converges in measure to
a minimizer of $F_{\gamma_w}$ in $\Gamma_q$. \end{lemma}
 {\bf Proof:} \
 Suppose $\bar{E} =
\lim_{\eps\rightarrow 0} E_\eps$. Evidently, $\bar{E}\in
\Lambda_q$. Suppose there exists $E_0\in \Lambda_q$ for which
$F_{\gamma_w}(E_0) < F_{\gamma_w}(\bar{E})$. According
to [b], there exists a subsequence $\eps_j\rightarrow 0$ and
$E^{'}_{\eps_j}\in \Lambda^\eps_q$ for which
$\lim_{j\rightarrow\infty}F^{\eps_j}_{\gamma_{\eps_j}}(E^{'}_j)=
F_{\gamma_w}(E_0)$. Then
$$ F_{\gamma_w}(E_0) \geq \lim_{j\rightarrow\infty} F^{\eps_j}_{\gamma_{\eps_j}}(E_{\eps_j})\geq F_{\gamma_w}(\bar{E})$$
where the last inequality follows from [a]. This
 contradicts the
assumption that $E_0$ is a minimizer of
$F_{\gamma_w}$ on $\Lambda_q$. \ \ \ \ \ \
$\Box$ \vskip .3in\noindent {\bf Proof of
Theorem~\ref{proofmain.1}:}  \ We need to verify parts [a] and [b]
of Lemma~\ref{proofmain.3}. To prove [a], consider \be
F_{\gamma_w}(E_0) - F^\eps_{\gamma_\eps}(E_\eps) = \left[
F_{\gamma_w}(E_0) - F^\eps_{\gamma_\eps}(E_0\cap\Omega_\eps)
\right] + \left[ F^\eps_{\gamma_\eps}(E_0\cap\Omega_\eps) -
F^\eps_{\gamma_\eps}(E_\eps)\right] \equiv (A) +
(B)\label{prfm}\ee For $\kappa>0$, define $$\Omega(\kappa) =
\left\{ x\in\Omega \ ; \  dist(x, \Gamma) < \kappa\right\}$$ By
assumption A2, there exists $\kappa = \kappa(\eps)$ such that
$\Omega_\eps \supset \Omega - \Omega(\kappa(\eps))$, and
\be\lim_{\eps\rightarrow 0} \kappa(\eps) = 0\label{kto0}\ee Then
\be (A) \leq \int_{\Omega(\kappa)}|\nabla\phi_{E_0}| +
\left(\int_{\Gamma_\eps}\gamma_\eps - \int_\Gamma\gamma_w
\right)\phi_{E_0} d {\cal H}_{n-1}\label{eee}\ee The second term
of (\ref{eee}) converges to $0$ by A3.  Using (\ref{kto0}) and Lemma \ref{main.2} we obtain that $(A)= o(1)$.
\par
To estimate (B),
$$F^\eps_{\gamma_\eps}(E_0\cap\Omega_\eps) - F^\eps_{\gamma_\eps}(E_\eps)
\leq
\int_{\Omega - \Omega(\kappa)}|\nabla \phi_{E_0}|
- \int_{\Omega - \Omega(\kappa)}|\nabla \phi_{E_\eps}|
+ \int_{\Omega(\kappa)}|\nabla\phi_{E_0}|
- \int_{\Omega(\kappa)\cap\Omega_\eps}|\nabla\phi_{E_\eps}|$$
\be + \int_{\Gamma_\eps}\left| \phi_{E_0} - \phi_{E_\eps}\right|
d {\cal H}_{n-1}\label{ddd}\ee
where we used $\gamma_\eps\leq \gamma_c< 1$ (A.4).
By (\ref{sspe}) applied to the domain $\Omega(\kappa)\cap\Omega_\eps$
we have
$$\int_{\Gamma_\eps} \left|\phi_{E_\eps} - \phi_{E_0}\right|
 d{\cal H}_{n-1} \leq
 \left[ \int_{\Omega(\kappa)\cap\Omega_\eps}\left|
\nabla \phi_{E_\eps}\right| + \int_{\Omega(\kappa)\cap\Omega_\eps}\left|
\nabla \phi_{E_0}\right|\right] +
\beta(\kappa)
 \int_{\Omega(\kappa)\cap\Omega_\eps}
\left|\phi_{E_\eps} - \phi_{E_0}\right|$$ where $\beta(\kappa)$ is
independent of $\eps$ (c.f. the remark below (\ref{ssspe}) in
section~\ref{main}). Hence \be (B) \leq \left[ \int_{\Omega -
\Omega(\kappa)}|\nabla \phi_{E_0}| - \int_{\Omega -
\Omega(\kappa)}|\nabla \phi_{E_\eps}|\right] +
2\int_{\Omega(\kappa)}|\nabla\phi_{E_0}| + \beta(\kappa)
 \int_{\Omega(\kappa)\cap\Omega_\eps}
\left|\phi_{E_\eps} - \phi_{E_0}\right|\label{hhh}\ee Fixing
$\kappa$ and letting $\eps\rightarrow 0$, the first and last terms
of (\ref{hhh}) has a nonpositive limit by the compactness Theorem
(section~\ref{review}). Now, we choose $\kappa = \kappa(\eps)$ and
use (\ref{kto0}) for the second term. This completes the proof of
assumption [a] of Lemma~\ref{proofmain.2}.
\par
The proof of part [b] is rather easy. As a first candidate to
$E_\eps^{'}=E_0\cap\Omega_\eps\in BV(\Omega_\eps)$. The second term in (\ref{prfm}) is identically zero while the first term is estimated as in  (\ref{eee}).
Since $E_0\cap\Omega_\eps\in BV(\Omega_\eps)$  does not satisfy
the volume constraint, we need to compensate the volume lost
$vol(E_0- \Omega_\eps) = {\cal O}(\eps)$. To do this, fix
$\delta > 0$. We can evidently find a ball
$B_\eps\subset\Omega-\Omega_\kappa$ of radius $\leq\delta$ such
that $vol(E_0\cap\Omega_\eps \cup B_\eps) = v$, where $\delta$
and $\kappa$ held fixed and $\eps$ sufficiently small. Then, by
Lemma~\ref{main}.2 with $\Omega = \Omega_\eps$, $\Omega_1=
B_\eps$, $\Omega_2 = \Omega_\eps - B_\eps$ and $\phi_i$ the
characteristic functions of $E^{'}_\eps \equiv
E_0\cap\Omega_\eps \cup B_\eps\in \Lambda^\eps_v$ restricted to
$\Omega_1$ and $\Omega_2$, respectively, we obtain
$$\int_{\Omega_\eps} \left|\nabla\phi_{E^{'}_\eps}\right| = 0 +
\int_{\Omega_2}\left|\nabla\phi_2\right| + \int_{\partial B_\eps}
d {\cal H}_{n-1} \leq
\int_{\Omega_\eps}\left|\nabla\phi_{E_\eps^{'}}\right| + {\cal
O}\left(\delta^{n-1}\right)$$ while the trace of $E^{'}_\eps$ on
$\Gamma_\eps$ is evidently identical to the trace of
$E_0\cap\Omega_\eps$. Hence \be\lim_{\eps\rightarrow 0}
F^\eps_{\gamma_\eps}\left(E^{'}_\eps\right) \leq \lim_{\eps\rightarrow 0}
F^\eps_{\gamma_\eps}\left(E_\eps\cap\Omega_\eps\right)  =
F_{\gamma_w}(E_0)\label{www}\ee where the equality in
(\ref{www}) follows immediately from part [a]. \ \ \ \ \ \ $\Box$
\section{Partial wetting-post critical interfacial energy}\label{pw}
In this section we deal with the case were the interfacial energy $\gamma<1$ is a constant and condition (A4) is
violated, i.e.
\be\label{nv} 1>\gamma>\gamma_{c}:=\inf_{x\in\Gamma} \vec{n}_\eps(x)\cdot \vec{v}(x) \ . \ee
For simplicity, again, we concentrate on smooth domains
$\Omega\subset \R^2$. We shall take the perimeter of $\Omega$ to
be $1$. Let $s$ be an arc-length parametrization of $\Gamma =
\vec{k}(s), \ \ 0\leq s< 1$. Let $\vec{n}(s)$ be the outward
normal to $\Gamma$ at the point $\vec{k}(s)$. Thus $\vec{n}(s)\cdot\dot{\vec{k}}(s)=0$.
\par
We shall describe the perturbed domain $\Omega_\eps$ by the
following: Let $\zeta$ be a smooth, positive  function
on $\R$ which is $1-$ periodic, namely $\zeta(s+1)=\zeta(s)$ for any $s\in\R$. For $\eps=1/j$ we parameterize $\Gamma_\eps$ to by \be\vec{k}_\eps(s):=\vec{k}(s) - \eps\zeta(s/\eps)\vec{n}(s) \
\ \ ; \ \ \ 0\leq s < 1 \label{par}\ee The domain
$\Omega_{\eps}$ is defined naturally as the interior of
$\Gamma_{\eps}$.
\par
By this definition, $\vec{n}(s)$ is perpendicular to $\vec{k}^{'}(s)$. Scaling $\bar{s}=s/\eps$ we also get to leading order
$$ \vec{k}_\eps^{'}(s)=\vec{k}^{'}(\eps\bar{s})-\zeta^{'}(\bar{s})\vec{n}(\eps\bar{s}) \ .  $$
Since $\vec{k}^{'}$ and $\vec{n}$ are orthonormal
   we obtain that $|\vec{k}_j^{'}|=\sqrt{1+|\zeta^{'}|^2}$ so the normal vector
$$ \frac{\vec{k}_j^{'}}{|\vec{k}_j^{'}|}= \frac{\vec{k}^{'}-\zeta^{'}\vec{n}}{\sqrt{1+|\zeta^{'}|^2}}$$
is perpendicular to the normal $\vec{n}_{\eps}$ of $\Gamma_{\eps}$. Thus
$$ \vec{n}_{\eps}\cdot\vec{n} = \frac{\vec{k}_j^{'}}{|\vec{k}_j^{'}|}\cdot \vec{k}^{'}= \frac{1}{\sqrt{1+|\zeta^{'}|^2} }\ . $$
By (\ref{nv}), identifying $\vec{n}$ with $\vec{v}$ in the limit $\eps=0$ we pose the condition
$$ \gamma>\gamma_c\equiv \sup_{s\in[0,1]} \frac{1}{\sqrt{1+|\zeta^{'}(s)|^2}}\ . $$
We also observe that the average roughness of this family $\Omega_{\eps}$ is
\be\label{rp}
r=\int_0^1\sqrt{1 + \left|\zeta^{'}\right|^2} \ .  \ee
To make things somewhat simpler, let us assume, in addition, that $\zeta$
is an even function which is monotone on the semi-period $[0,1/2]$. Let
$s=h(y)$ be the inverse of $\zeta$ on this interval. The function $h$ is
defined on the interval $[0,Y]$ where $Y=\max\zeta$ with $h(0)=1/2$,
$h(Y)=0$ (cf. Fig [2]). In terms of $Y$ we recover
\be\label{hcon}\gamma_c=\inf_{0\leq y\leq Y}\frac{|h^{'}(y)|}{\sqrt{1+(h^{'}(y))^2}} \ \ \ ; \ \ \ r= \int_0^Y\sqrt{1 + \left|h^{'}\right|^2}  \ . \ee
Define
$$g_\gamma(y) = h(y) + \gamma \int_0^y\sqrt{1 + \left|h^{'}(y)\right|^2} \ . $$
\begin{figure}
 \centering
\includegraphics[height=8.cm, width=12.cm]{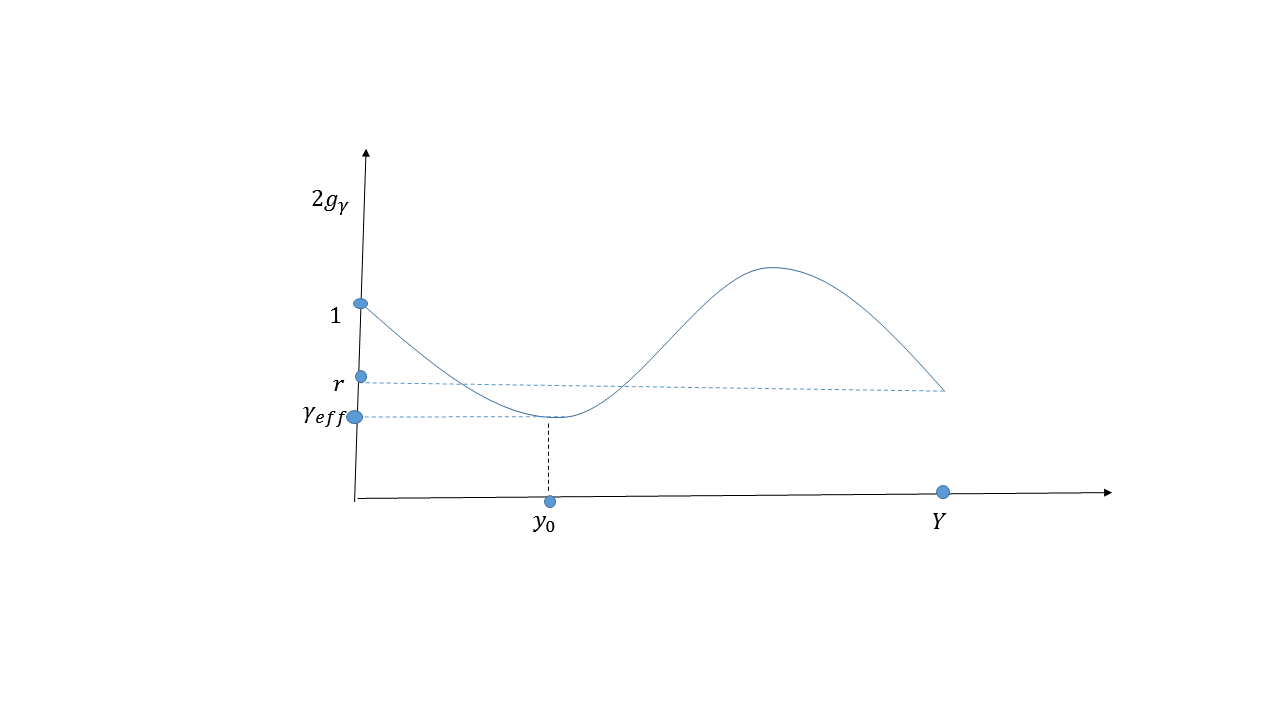}\\
\caption{Plot of $2g_\gamma$ vs. $y$. Here $1>r>\gamma_{eff}$.}
\label{fig5}
\end{figure}
Note that $h^{'}(0)=h^{'}(Y)=-\infty$. Since $\gamma<1$ it follows that $g_\gamma$ is decreasing near $y=0$ and $y=Y$. If $\gamma<\gamma_c$ then by (\ref{hcon}) we find that $g_\gamma$ is decreasing on the whole interval $[0,Y]$. If, however,  we assume $1>\gamma>\gamma_c$ then we obtain that there is an interval in $[0,Y]$ in which $g_\gamma$ is increasing. In that case, let
$$\gamma_{eff}\equiv  2 \inf_{y\in [0,Y]} g_\gamma(y)=  2 g_\gamma(y_0)\eqno{(A5)}$$
and assume $g$ is monotone decreasing on the interval $[0, y_0]$.
Note that by (\ref{hcon}) $2g_\gamma(Y)=r\gamma$  while $2g_\gamma(0)=1$ by definition.\footnote{Note that this definition is equivalent to (\ref{gammaeff}).} Hence $\gamma_{eff}
< \min\{1, r\gamma\}$.
In particular,
 $\gamma_{eff} < 1$.
\par
Consider the domain
\be D= (x,y); y_0\leq y \leq Y, \ -h(y) \leq x\leq h(y)\label{Ddef}\ee
$$\partial D = \Gamma_1 \cup \Gamma_2$$
where
$$ \Gamma_1 = \{ -h(y_0) \leq x\leq h(y_0) \} \ ,  \ y=y_0 \ \ \ \ ; \ \ \
\ \Gamma_2=\partial D - \Gamma_1$$

\begin{figure}
 \centering
\includegraphics[height=5.cm, width=10.cm]{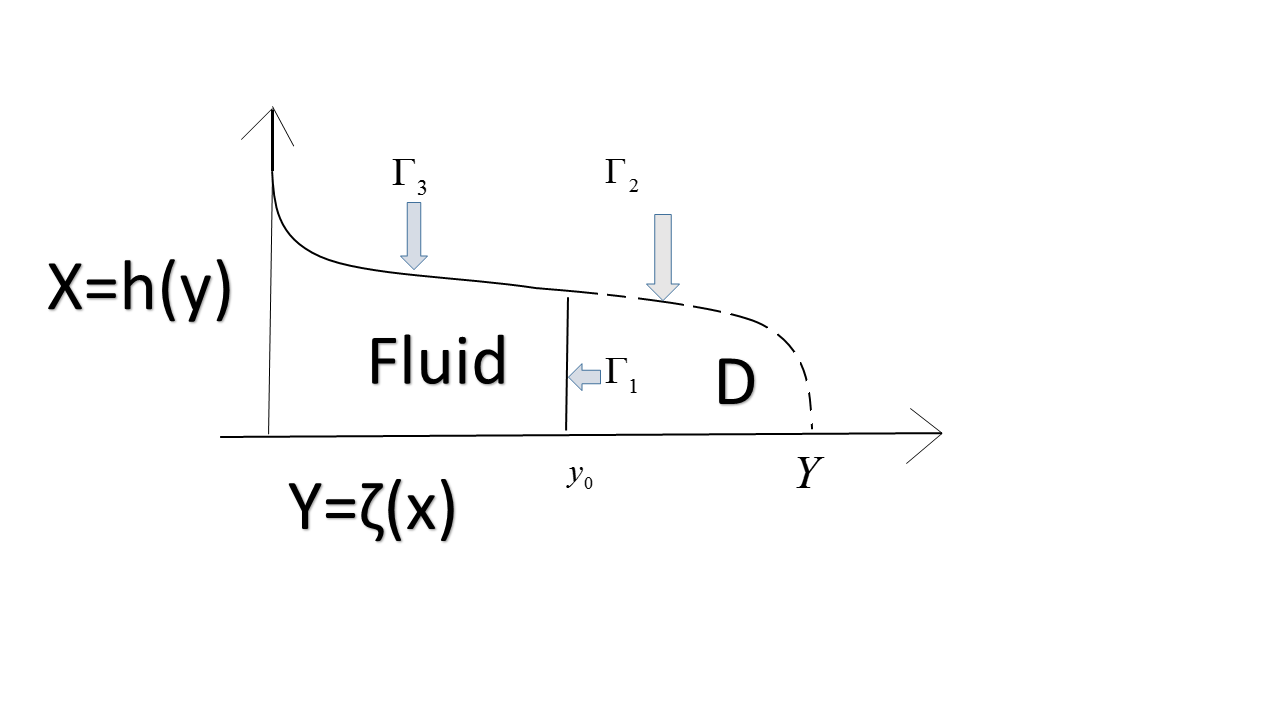}\\
\caption{The inaccessibility domain D}
\end{figure}
Domain $D$ is called {\em unreachable} if
 \be\label{unreachableD}F_D(A):= \int_D\left|\nabla\phi_A\right| -
\int_{\Gamma_1}\phi_A + \gamma\int_{\Gamma_2}\phi_A \geq 0\ \ \ \
\forall A\in BV(D)\ee
\par\noindent
To make this condition more explicit, we pose the following
\par\noindent
\vskip .2in\noindent {\bf Proposition~\ref{pw}.1}: \ {\it Suppose
there exists a vector-field $(w_1, w_2):= \vec{w}\in
C^1\left(\overline{D}
; \R^2\right)$ with the following properties:
\begin{description}
\item{a.} \ $\sup_D\left|\vec{w}\right| \leq 1$
\item{b.} \ $\nabla\cdot \vec{w} \geq 0$ on $D$
\item{c.} \ $\vec{w}\cdot \vec{\nu} \leq \gamma$ on $\Gamma_2$ where
$\vec{\nu}$
is the outer normal to $\partial D$.
\item{d.} \ $w_1=1$ on $\Gamma_1$ (i.e $\vec{w}\cdot\vec{\nu} = -1$
on $\Gamma_1$).
\end{description}
Then $D$ is unreachable.}
\par\noindent
{\bf Proof:} \  By the divergence theorem applied to a BV-function
$\phi\geq 0$ we have: $$0\leq  \int_D \phi\nabla\cdot\vec{w} =
-\int_D\nabla\phi\cdot\vec{w} + \int_{\partial D}\phi\vec{w}\cdot
\nu \leq \int_D|\nabla\phi| - \int_{\Gamma_1} \phi +
\gamma\int_{\Gamma_2}\phi$$ where the last inequality follows from
(c) and (d). \ \ \ $\Box$
\par
Proposition~\ref{pw}.1 is close to a criterion introduced by Finn \cite{[F]}, p. 145.
Note that (c) and (d) are consistent with (b) by the divergence theorem
and  (A5) via:
$$\int_D\nabla\cdot\vec{w}=
\int_{\partial D} \vec{w}\cdot\vec{\nu} \geq 2(g_\gamma(Y)-g_\gamma(y_0)) \geq 0$$
We now derive an explicit sufficient condition for $D$ to be unreachable. If
$\int_{y_0}^Y h^{-1} \leq 1$ set $y_1=Y$, else determine $y_1$ from the
condition
$$ \int_{y_0}^{y_1}\frac{1}{h} =1$$
\par\noindent{\bf Lemma~\ref{pw}.2}:\ {\it Suppose
$$-h^{'}(y)\sqrt{ 1 - \left( \int_{y_0}^y\frac{1}{h(y)}\right)^2}
+ \int_{y_0}^y\frac{1}{h} \leq \gamma\sqrt{ 1 + \left(h^{'}(y)\right)^2}$$
for $y_0\leq y\leq y_1$, and
$$1 \leq \gamma\sqrt{ 1 + \left(h^{'}(y)\right)^2}$$
for $y_1\leq y\leq Y$. Then $D$ is unreachable.}

\par
A more general (but less explicit) condition for unreachable $D$ is given by:
\par\noindent{\bf Lemma~\ref{pw}.3}:\ {\it Suppose there exists
a pair of functions $\sigma$, $\beta$ on the interval $[y_0, Y]$
such that  the following hold on this interval:
\begin{description}
\item{a.} \ $\sigma^2 + h^2\beta^2 \leq 1$
\item{b.} \ $\sigma^{'} + \beta\geq 0$
\item{c.}\ $-h^{'}\sigma + h \beta\leq \gamma
\sqrt{1 + \left( h^{'}\right)^2}$
\item{d.} \ $\sigma(y_0) =1$
\end{description}
Then $D$ in unreachable.}
\par
Lemma~\ref{pw}.3 follows from Proposition~\ref{pw}.1 where the
vectorfield $\vec{w}$ is given by:
$$\vec{w} = x \beta(y){\bf e}_x +\sigma(y){\bf e}_y $$
here ${\bf e}_x$, ${\bf e}_y$ are the vector coordinates
in the $x,y$ directions, respectively. One can check easily that
conditions (a-d) of Proposition~\ref{pw}.1 correspond to those of
Lemma~\ref{pw}.3.
\par

To obtain the proof of Lemma~\ref{pw}.2, use (a) to define
$\beta = h^{-1}\sqrt{1-\sigma^2}$ and substitute in (b). This gives the
differential inequality $\sigma^{'} + h^{-1}\sqrt{ 1-\sigma^2}\geq0$.
A solution of this inequality is given by:
$$ \sigma(y) = \sqrt{\left[1 - \left(\int_{y_0}^y \frac{1}{h}
\right)^2\right]_+}
 $$
Now substitute this $\sigma$ in condition (c) of Lemma~\ref{pw}.3 to obtain
the condition of Lemma~\ref{pw}.2.
\par
We are now in a position to state the main result for the
partial-wetting case:
\begin{theorem}\label{pw.1}  $\Omega_\eps\subset \R^2$ are a family of smooth domains
 parameterized by (\ref{par}),
where $\eps=1/j$, $\zeta$ is an even, nonnegative 1-periodic, smooth function
which is monotone on its semi-period. Assume $1>\gamma>\gamma_c$ {\em is a  constant}.
Assume $D$ determined by (\ref{Ddef}) is unreachable. Let $E_{\eps}\subset\Omega_\eps$ be a minimizer of $\Feps$ under a volume constraint. Then the limit of  $E_\eps$ converges, as $j\rightarrow\infty$, to a minimizer $E_0$ of  $F_{\gamma_{eff}}$ in the limit domain $\Omega$ under the same volume constraint, where  $\gamma_{eff}$ given by (A5).
\end{theorem}

\noindent{\bf Proof:} \
Let ${\cal O}_{\eps}$ be the interior domain of
$$\vec{k}(s) - \eps y_0\vec{n}(s) \ \ \ ; \ \ \ 0\leq s< 1$$
where $y_0$ defined in (A5)
and $\hat{\Omega}_{\eps}$ be given by
$\Omega_{\eps}\cap{\cal O}_{\eps}$.
Set $\hat{\Gamma}_{\eps}\equiv\partial\hat{\Omega}_{\eps}=
\hat{\Gamma}_{\eps}^{(1)} \cup \hat{\Gamma}_{\eps}^{(3)}$ where
\be\label{Gammahat}\hat{\Gamma}_{\eps}^{(3)} = \partial\Omega_{\eps}\cap
{\cal O}_{\eps} \ \ \ ; \ \ \
\hat{\Gamma}_{\eps}^{(1)}
\equiv
\partial\hat{\Omega}_{\eps}-
\hat{\Gamma}_{\eps}^{(1)}\ \ \text{and} \ \  \hat{\Gamma}_{\eps}^{(2)} =\partial\Omega_\eps-\hat{\Gamma}_{\eps}^{(3)}\ .  \ee
 Let $0<\delta_n<1-\gamma$ and set
$$\gamma^{\delta_n}_\eps(x): = \left\{\begin{array}{cc}
                        \gamma & \text{if} \ x\in \hat{\Gamma}_{\eps}^{(3)} \\
                       1-\delta_n &  \text{if} \ x\in \hat{\Gamma}_{\eps}^{(1)}
                     \end{array}\right.
                    $$
We claim that $\hat{\Omega}_{\eps}$ and $\gamma_\eps$ so defined
satisfy the assumptions
of Theorem~\ref{proofmain.1}. Evidently, $\hat{\Gamma}_{\eps}$
is Lipshitz and satisfies (A1-A2). By  (A5)
the roughness parameter of $\hat{\Omega}_{\eps}$ is
\be\label{gammaeffdelta}\gamma_{eff}^{\delta_n}:=2g_\gamma(y_0)/ \gamma-O({\delta_n})\  , \ee
so  condition (A3) is satisfied with $\gamma_{eff}^{\delta_n}$ replacing $\gamma_w$
(cf. Proposition~\ref{rough}.2). We only need to show condition (A4) for $\gamma_\eps$.

To see this, first note that the normal
$\vec{n}_\eps$ at any point of $\hat{\Gamma}^{(1)}_{\eps}$ is identical
(up to $O(\eps)$) to $\vec{v}=\vec{n}$ at this point, hence, for $\eps<<{\delta_n}$
$$ \vec{v}\cdot \vec{n}_\eps = 1-O(\eps) > \gamma^{\delta_n}_\eps(x)
\ \ \ \ \forall x\in \hat{\Gamma}^{(1)}_{\eps}$$
Now let $x\in \hat{\Gamma}^{(3)}_{\eps}$. If we blow-up the coordinate
system near this point by the $\eps$ scale and rotate the coordinate
system such that $\vec{n}$ coincide with the  $y$ coordinate vector
${\bf e}_y$ at this point, we get $\vec{n}_\eps$ in the direction
(up to $O(\eps)$ error)
of the normal to the graph of $\zeta$ at the corresponding point. Hence
\be\vec{v}\cdot \vec{n}_\eps = \left[ 1 +\left(\zeta^{'}(s)\right)^2
\right]^{-1/2} + O(\eps)=
 \frac{\left| h^{'}(y)\right|}{\sqrt{1 +\left(h^{'}(y)\right)^2}}
 + O(\eps)\label{vdotneps}\ee
 \begin{figure}
 \centering
\includegraphics[height=6.cm, width=8.cm]{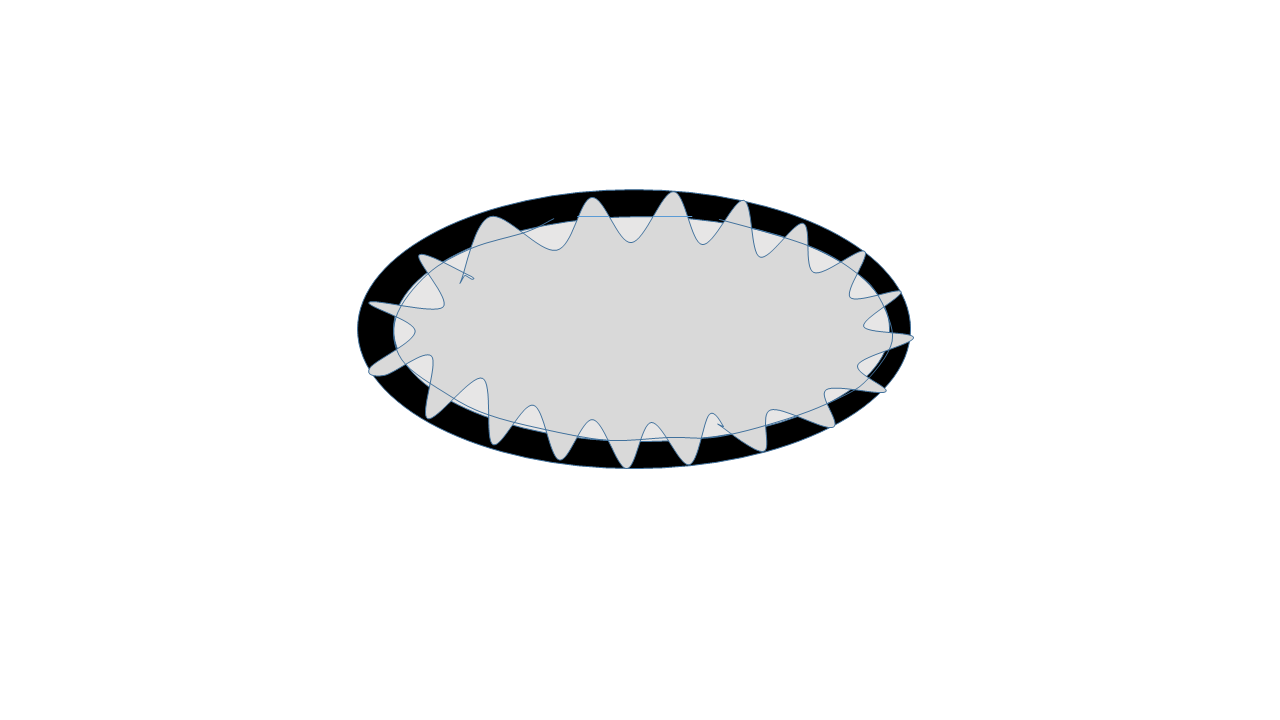}\\
\caption{$\Omega-{\cal O}_\eps$ \ - \  black domain. $\hat{\Omega}_\eps$-light gray}
\end{figure}
Now, $\hat{\Gamma}^{(3)}_{\eps}$ corresponds to $y\in [0, y_0]$.
By assumption, $g_\gamma$ is monotone non-increasing on this interval
(see Fig. \ref{fig5}.  This, in fact, is implied from $D$ being unreachable), hence:
$$ g^{'}(y) = h^{'}(y) + \gamma\sqrt{1 + \left| h^{'}(y)\right|^2}\leq 0$$
for $y\in [0, y_0]$. Using this in (\ref{vdotneps}) we obtain that
condition (A4) is satisfied on $\hat{\Gamma}^{(3)}_{\eps}$ as well.
\par
We can now repeat the proof of Theorem~\ref{proofmain.1} line by line,
provided we replace $F^{\eps}_{\gamma_\eps}$ by the free energy
$\hat{F}^{\eps}_{\gamma^{\delta_n}_\eps}$
corresponding to the domain $\hat{\Omega}_{\eps}$ and $\gamma_\eps^{\delta_n}$. Let $E_\eps^{\delta_n}$ be the minimizers of $\hat{F}^{\eps}_{\gamma_\eps}$. Then by Theorem ~\ref{proofmain.1}, there exists a subsequence of $E_\eps^{\delta_n}$ converging to a minimizer $E_0^{\delta_n}$ of  $F_{\gamma_{eff}^{\delta_n}}$, along which
\be\label{limitn} \lim_{\eps\rightarrow 0}\hat{F}^{\eps}_{\gamma^{\delta_n}_\eps}(E_\eps^{\delta_n})=F_{\gamma_{eff}^{\delta_n}}(E_0^{\delta_n}) \ . \ee
Let $\delta_n\rightarrow 0$. By (\ref{gammaeffdelta},\ref{limitn}) we can obtain another subsequence $\eps_n\rightarrow 0$  and
$E_{\eps_n}^{\delta_n}\rightarrow E_0$ for which
\be \label{hatFedelta}\lim_{n\rightarrow \infty}\hat{F}^{\eps_n}_{\gamma^{\delta_n}_\eps}(E_{\eps_n}^{\delta_n})=F_{\gamma_{eff}}(E_0) \ . \ee
and $E_0$ is a minimizer of $F_{\gamma_{eff}}$.

We now prove that for any $\eta>0$
\be  \hat{F}^{\eps_n}_{\gamma^{\delta_n}}
\left(E_{\eps_n}\cap\hat{\Omega}_{\eps_n}\right)- F^{\eps_n}_{\gamma}\left(E_{\eps_n}\right)<\eta
\label{error}\ee
for sufficiently large $n$. Here
 $E_{\eps_n}\subset\Omega_{\eps_n}$ is a minimizer of $F^{\eps_n}_\gamma$. This estimate, together with (\ref{hatFedelta}), implies
 $$F^{\eps_n}_{\gamma}(E_{\eps})\geq F_{\gamma_{eff}}(E_0)-\eta$$
 for sufficiently large $n$.  Since $E_{\eps_n}^{\delta_n}\subset\hat{\Omega}_{\eps_n}\subset\Omega_{\eps_n}$, we obtain that $E_0$ is the limit of   a recovery sequence in the sense of Lemma  \ref{proofmain.3}.

 By Lemma~\ref{main.2} we may rewrite
(\ref{error}) as
\be-\int_{\Omega_{\eps} - \hat{\Omega}_{\eps}} \left|
\nabla \phi_{E_{\eps}}\right|+ (1-\delta_n)\int_{\hat{\Gamma}^{(1)}_{\eps}}
\phi_{E_{\eps}}- \gamma\int_{\hat{\Gamma}^{(2)}_{\eps}}
\phi_{E_{\eps}} \ , \label{error1}\ee
where we used the fact that both functionals attribute the same trace, $\gamma$, to $\hat{\Gamma}_\eps^{(3)}\subset\partial\Omega_\eps$, and
$\partial\Omega_\eps=\hat{\Gamma}^{(2)}_\eps\cup \hat{\Gamma}^{(3)}_\eps$ (\ref{Gammahat}).
Observe that $\Omega_{\eps}- \hat{\Omega}_{\eps}$ can be written as
the union of $m=1/[\eps]$ cells $D^{(i)}_{\eps}$, $i=1, \ldots m$.
Let
$A^{(i)}_{\eps} = E_{\eps}\cap D^{(i)}_{\eps}$. Then (\ref{error1})
is rewritten as
$$-\sum_{i=1}^n F_{D^{(i)}_{\eps}} \left(
A^{(i)}_{\eps}\right)$$
where
$$F_{D^{(i)}_{\eps}}\left(A\right)\equiv \int_{D^{(i)}_{\eps}}
\left|\nabla\phi_A\right| - (1-\delta_n)\int_{\Gamma^{1, (i)}_{\eps}}\phi_A
+ \gamma\int_{\Gamma^{2,(i)}_{\eps}}\phi_A \geq 0\ \ \ \
\forall A\in BV(D^{(i)}_{\eps})$$
 and $\Gamma^{\cdot, (i)}_{\eps}$ are the components of the boundary of
 $D^{(i)}_{\eps}$.
\par
We claim now that for each $\eta >0$ there exists $N$ such that
\be F_{D^{(i)}_{\eps}} \left(
A^{(i)}_{\eps}\right)\geq -
\eta O\left(1/m\right) \ \ \ ; \ \ \ \forall m>N \ \ ; \ \ \forall i=1, 2\ldots m
\label{fsubd}\ee
To see this,  rotate one of the cells $D^{(i)}_{\eps}$
so that the normal to $\Omega$ at the corresponding point is pointing
in the direction of ${\bf e}_y$ and expand the $x-y$ coordinates by:
$\{ x, y\} \rightarrow \{mx, my\}$, we obtain a
domain $mD^{(i)}_{1/m}$ which
is a smooth $\eps=1/m$ deformation of the domain $D$ defined in (\ref{Ddef}).
The corresponding $F_{D^{(i)}_{1/m}}$ is transformed into
$\eps \tilde{F}^{\eps}_D$ (recall $m=[1/\eps]$) where
$$ \tilde{F}^{\eps}_D(A) =
\int_D\left|\sigma^{(0)}_{\eps}\nabla\phi_A\right| -(1-\delta_n) \int_{\Gamma_1}
\sigma^{(1)}_{\eps}\phi_A
+ \gamma\int_{\Gamma_2}\sigma^{(2)}_{\eps}\phi_A $$
and $\sigma^{(\cdot)}_\eps$ are related to the
Jacobian of the above deformation.
Thus
\be\left|\sigma^{(k)}_{\eps} -1\right|_\infty <\eta ,\ \ \text{for} \ \eps\ \ \text{small enough}, \  \  k=0,1,2\label{sigma}\ee
 so
$$F_{D^{(i)}_{\eps}} \left(
A^{(i)}_{\eps}\right) = \eps\tilde{F}^{\eps}_D(A)\geq \eps\left(F_D(A) - \sup_{k=0,1,2}
\left|\sigma^{(k)}_{\eps} -1\right|_\infty \left(
\left|\nabla\phi_A\right|_1 + \int_{\Gamma^1 \cup \Gamma^2}
\left|\phi_A\right|\right)\right)$$
By the assumed (\ref{unreachableD}) and (\ref{sigma}) we have
(\ref{fsubd}), and the required estimate on (\ref{error}).
The rest of the proof goes exactly as the proof of Theorem~\ref{proofmain.1}.
\ \ \ \ \ \ \ $\Box$
%
%

\vskip 2.in\noindent
{\it Acknowledgment:} I wish to thank Prof. Avi Marmur for
introducing me to this subject, many years ago..


\vskip .5in

\begin{thebibliography}{1}
\bibitem{[CS]} R. Chebb and M. Sami Selim, {\it Capillary spreading of
liquid drops on solid surfaces}, J. Colloid Interface Sci, {\bf 195},
(1997), 66-76

\bibitem{[CPM]}
D. Cwickel,  Y.  Paz   and A.  Marmur,
{\it Contact angle measurement on rough surfaces: the missing link},
Surface Innovations,
doi = {10.1680/jsuin.17.00021},
URL = {
        https://doi.org/10.1680/jsuin.17.00021}



\bibitem{[D]} B.V, Derjaguim, C.R Acad. Sci. USSR {\bf 51}, (1956), 361
\bibitem{[E]} M.Emmer, {\it Esistenza, unicit$\grave{a}$ e regularit$\grave{a}$ nelle
superfici di equilibrio nei capillari}, Ann Univ. Ferrara Sez. VII
{\bf 18} (1973), 79-94
\bibitem{[F]} R, Finn, {\it Equilibrium Capillary Surfaces},
Springer-Verlag, (1986)


\bibitem{[G1]} E. Giutsi, {\it Minimal Surfaces and Functions of
Bounded Variation}. Birkh$\ddot{a}$user, Boston (1984)
\bibitem{[G]} R.J. Good, J. Amer. Chem. Soc. {\bf 74}, (1952) 5041

\bibitem{[G2]} E. Giutsi, {\it The equilibrium configuration of
liquid drops}, J. Reine Angew. Math. {\bf 321} (1981), 53-63


\bibitem{[HM]} C. Huh and S.G.Mason {\it Effects of surface roughness on wetting
(Theoretical)}, J. Colloid Interface Sci, {\bf 60}, (1977), 11-38
\bibitem{[JBGM]} B. Jan\'czuk, J.M Bruque, M.L. Gonza\'lez-Mart\'in
and J. Moreno Del Vozo
{\it Determination of components of Cassiterite surface free-energy from
contact angle measurements}, J. Colloid Interface Sci, {\bf 166} (1993)
209-222
\bibitem{[JD]} R.E Johanson and R.H Dettre,  {\it Contact angle
hysteresis},  Contact angle, Wettability and Adhesion, Adv. Chem. Ser.,
{\bf 43}, ACS, (1964), 136-144
\bibitem{[L]}  P.L Laplace, {\it Trait$\acute{e}$ de M$\acute{e}$caniqe
C$\acute{e}$leste;
suppl$\acute{e}$mes au Livre X}, 1805 and 1806 resp. in Euvres
Complete Vol. 4. Gauthier-Villars, Paris

 \bibitem{[AMAR]}
        A. Marmur,  {\it Wetting of Hydrophobic Rough Surfaces: To be heterogeneous or not to be}, . Langmuir. 19 (20): 8343-8348, (2003)

\bibitem{[M1]} A. Marmur, {\it Equilibrium contact angles: Theory and
measurements}, Colloids and Surfaces A., {\bf 116},  (1996), 55-61
\bibitem{[M2]} A. Marmur, {\it Thermodynamic effects of contact angle
hysteresis}, Adv. Colloids and Interface Science, {\bf 50},
(1994), 121-141
\bibitem{[M3]} A. Marmur {\it Contact angle hysteresis on heterogeneous
smooth surfaces}, J. Colloid Interface Sci, {\bf 186}, (1994), 40-46
\bibitem{[MM]} G. Mason and N.R. Morrow, {\it Effect of contact angle on
capillary displacement curvature in pore throats formed by spheres},
J. Colloid Interface Sci, {\bf 168}, (1994), 130-141
\bibitem{[NVZ]} Y.V. Naidich, R.P Voitovich and V.V Zabuga {\it Wetting and
spreading in heterogeneous solid surface- Metal melt systems},
J. Colloid Interface Sci, {\bf 174},
(1995), 104-111
\bibitem{[NC]} A. Novic-Cohen, {\it A singular minimization problem for
 droplet profiles}, Euro. Jn. Applied Mathematics, {\bf 4},
  (1993),
 399-418

\bibitem{[SG]} L.W. Schwartz and S. Garoff, {\it Contact angle hysteresis and the
shape of the three-phase line} , J. Colloid Interface Sci, {\bf 106}, (1985),
422-437, See also {\it contact angle hysteresis on heterogeneous surfaces},
Langmuir, {\bf 1985}, I, 219-230

 \bibitem{[S]} A. Sharma {\it Equilibrium contact angles and film thickness
in the apolar and polar systems: Role of intermolecular interactions
in coexistence of drops and thin films}, Langmuir, {\bf 9}, (1993),
3580-3586

\bibitem{[SB]} R. Shuttleworth and G.L.J Bailey, Discuss Faraday Soc. {\bf 3}, (1948)



\bibitem{[TLB]} J. Troger, K. Lunkwitz and W. Burger, {\it Determination of the
surface tension of microporus membrance using contact angle
measurements}, Jn. Colloid Interface Sci, {\bf 199}, (1997),
281-286



\bibitem{[WN]} R.N Wenzel, {\it Resistance of solid surfaces to
wetting by water}, Industrial and Engineering Chemistry, {\bf 28},
(1936) 988-994
\bibitem{[WM]}
G. Wolansky and A. Marmur, {\it Actual contact angle on a heterogeneous
rough surfaces in three dimensions}, Langmuir, {\bf 14}, (1998)
5292-5297
\bibitem{[Y]} T. Young, {\it An essay on
the cohesion of fluid}s, In Miscellaneous Works, (G. Peacock,
ed.) ,{\bf I}, John Murray, London, (1855), 418-453


\end{thebibliography}
\end{document}